\RequirePackage{tikz}
\documentclass[pdflatex,sn-basic]{sn-jnl}

\jyear{2021}%
\theoremstyle{thmstyleone}%
\theoremstyle{thmstyletwo}%

\theoremstyle{thmstylethree}%
\usepackage{lineno,hyperref}

\usepackage{natbib}
\usepackage{dblfloatfix}
\usepackage{enumitem}
\usepackage{comment}

\usepackage{algorithm}
\usepackage{algpseudocode}
\usepackage{graphicx}
\usepackage{multirow}

\usepackage{amsmath,amssymb,amsfonts}
\usepackage{textcomp}
\usepackage{xcolor}

\usepackage[T1]{fontenc}
\usepackage[utf8]{inputenc}
\usepackage{booktabs}
\usepackage{graphicx}
\usepackage{booktabs}
\usepackage{graphicx}

\usepackage{pgfplots} 
\pgfplotsset{compat=newest}
\usepgfplotslibrary{groupplots}
\usepgfplotslibrary{dateplot}
\usepackage{tikz}
\usepackage{xlop}
\begin{document}

\title[\footnotesize{Attention-Based Recurrent Neural Network For Automatic Behavior Laying Hen Recognition}]{Attention-Based Recurrent Neural Network For Automatic Behavior Laying Hen Recognition}


\author*[1]{\fnm{Fréjus A. A.} \sur{Laleye}}\email{frejus.laleye@opscidia.com}

\author[2]{\fnm{Mikaël A.} \sur{Mousse}}\email{mikael.mousse@univ-parakou.bj}

\affil*[1]{\orgname{Opscidia}, \orgaddress{\city{Paris}, \country{France}}}

\affil[2]{\orgdiv{Université de Parakou}, \orgname{Institut Universitaire de Technologie}, \orgaddress{\city{Parakou},  \country{Bénin}}}


\abstract{One of the interests of modern poultry farming is the vocalization of laying hens which contain very useful information on health behavior. This information is used as health and well-being indicators that help breeders better monitor laying hens, which involves early detection of problems for rapid and more effective intervention. In this work, we focus on the sound analysis for the recognition of the types of calls of the laying hens in order to propose a robust system of characterization of their behavior for a better monitoring. To do this, we first collected and annotated laying hen call signals, then designed an optimal acoustic characterization based on the combination of time and frequency domain features. We then used these features to build the  multi-label classification models based on recurrent neural network to assign a semantic class to the vocalization that characterize the laying hen behavior. The results show an overall performance with our model based on the combination of time and frequency domain features that obtained the highest F1-score (F1=$92.75$) with a gain of $17\%$ on the models using the frequency domain features and of $8\%$ on the compared approaches from the litterature.}




\maketitle

\section{Introduction}

Animal vocalisations are associated with different animal responses and can be used as useful indicators of the state of animal welfare. They are information about animal behavior allowing to determine the needs of the animals, providing
personalized and optimal attention for the benefit of the production \citep{doi:10.1080/14488388.2009.11464794, BARDELI20101524}.
There are two types of poultry farming which coexist: traditional poultry farming and modern poultry farming which is recent and is gaining more and more importance. Unlike traditional poultry farming, which is less demanding, the establishment of modern poultry farming is subject to investment no less negligible and, requires rigorous conduct. Well conducted, modern poultry farming constitutes a source of unquestionable fortune for Poultry Farmers. Indeed, with the increase in demand for poultry products in the market and the presence of other factors such as consumers demanding more transparency in reporting on the welfare, environmental impact and safety of poultry products, it is essential to think on a rationalization in the treatment of animals. 

This work was carried out on farms in the cities of the Republic of Benin, a West African country where the breeding of laying hens is common and practiced by many families. The poultry farming sector is booming and growing on a large scale in the country.
Commercial poultry farming is dominated by the breeding of laying hens for production of eggs which then experienced a spectacular boom in recent years becoming for some families the primary activity and the only source of income. However, the laying hen breeding is not an easy activity. With advances in technology, commercial farming has become modern and can use cameras, microphones, and sensors in
proximity to animals to take the place of farmers’ eyes and ears in monitoring animal health and welfare effectively \citep{DU2020473}.

In the context of poultry farming, laying hens always need more attention for better productivity in the sense that each of their vocalizations has a physiological meaning linked to their state of health. In this paper, we propose a recognition system to recognize and interpret the laying hen vocalizations, based on the sound calls that we collected ourselves, for helping the farmers in their breeding. 
The main contributions of this work are the following. First,  we built an annotated dataset\footnote{Location will be public after review} of different laying hen calls. These are the laying hen vocalizations that were recorded under different environment conditions with specific sensors. Second, we carried out a study of the acoustic features of laying hen vocalizations which led us to propose in a significant way the optimal representations of each vocalization type. Finally, based on the rich representations of vocalizations, our work has aimed to explore and develop an optimal recognition multi-label model for classifying the laying hen behaviours.\\
The rest of this article is organized as follows. Section \ref{related} presents some related works. Section \ref{task} introduces and reformulates the task of behavior laying hen recognition we addressed in this work. Section \ref{dataset} presents the data we collected for building our classifiers. Section \ref{approach} gives an overview of the approach we proposed. Experimental setup and result analysis is given in Section \ref{results}.
\section{Related Works}
\label{related}
Usually, farmers rely on their observations and experiences to manage their animals. In commercial farming it is complicated to monitor the animals and nowadays we see the use of the sensors (cameras and microphones) to help the farmer in his management. 
Several research works have investigated the use of sensors for the recognition of animal states. Liz et al.\cite{liz2020617625} and Garcia et al.\cite{GARCIA2020105826} have provided a comprehensive review of techniques for monitoring the behavior and welfare of broilers and laying hens. 
Du et al.\cite{DU2020473} proposed  a hen vocalisation detection method based on machine learning to assess their thermal comfort condition.  They used temporal and spectral features and developed a support vector machine (SVM) based classifier to estimate thermal confort of the chicks. Pluk et al.\cite{pluk201029} also used sound analysis to model the water intake of the broilers as response to the light input. The results of their experimentations have shown a high potential of using water intake for the monitoring of broiler production over a complete growing period. 
Chelotti et al. \cite{CHELOTTI201664} proposed a method for an automated detection and classification of
bite and chew activity has been used as means of feed intake estimation  in cattle. 
Other research works use sound analysis to interepret the behaviour of pigs. Thus, it has been demonstrated that respiratory diseases detected in pigs have an influence on their cough \citep{Ferrari200810511055,hirtum20041015}. 
Vandermeulen et al. \cite{VANDERMEULEN201615} proposed a method for the early recognition of bovine respiratory diseases in calves through automated continuous monitoring of cough sounds.\\

To analyze the vocalization sounds, several research works have based their approach on the widely Mel-Frequency Cepstral coefficient features that have competitive results in the litterature \citep{LEE200693,SHARAN201724,Pattanayak2021PitchrobustAF}.  

After the feature extraction, it is necessary to have a classification system capable of recognizing the state of the animal. To this end, several machine learning approaches have been proposed such as Support
Vector Data Descriptions (SVDD) \citep{chung2013automatic,chung2013automatica},  Support Vector Machines (SVM) \citep{bishop2017sound,BANAKAR2016744,Lee2015StressDA}, AdaBoost.M1 \citep{lee2014formant}, Deep Neural Networks (DNN)-pattern-recognition algorithms \citep{DBLP:journals/sensors/CowtonKPB18, Doulgerakis2019AnAW, 9013548} and Convolutional Neural Networks (CNN) \citep{doi:10.1098/rsif.2021.0921}. In \citep{DBLP:journals/sensors/CowtonKPB18}, the authors used the Recurrent Neural Networks (RNN) to describe and evaluate a methodology for the early warning
of anomalies pertaining to respiratory disease in growing pigs, based on environmental factors. 

In this paper, we propose a deep learning approach also based on RNN but unlike the work proposed in \cite{DBLP:journals/sensors/CowtonKPB18}, we introduced in our model the attention mechanism to recognize the behavior of laying hens. To do this, we based our approach on the acoustic analysis of vocalizations that identify behaviors. One of the challenges of our approach is the recognition of semantically close vocalizations that may share acoustic features. 
In \cite{Lee2015StressDA}, the authors tackled the challenge of laying hen vocalization detection. Their approach involved a hierarchical combination of three binary SVM classifiers, which were used to classify stress into two distinct semantic classes: physical stress and mental stress. In \citep{doi:10.1098/rsif.2021.0921}, the authors have presented compelling evidence that deep Convolutional Neural Network (CNN) models are highly effective in identifying distress vocalizations. Specifically, they fine-tuned a pre-trained CNN model using their dataset of chicken vocalizations, resulting in a notable boost in performance for identifying distress vocalizations. 
Despite yielding impressive accuracy scores and irrespective of the deep learning models utilized, these two approaches are not directly applicable to our work. We addressed a task that involves a more diverse range of vocalizations from laying hens, which may exhibit considerable variability and overlap. While these approaches have shown promise in previous studies, our work presents unique challenges that require a more tailored approach.


\section{Task Formalization}
\label{task}
In this work, we considered the problem of the automatic recognition of laying hen calls as a problem of identification and classification of sounds correlated with the environment in which the sounds are emitted. The aim is to assign a semantic state to the input audio stream that characterizes the condition in which the laying hen is – for example food calls, distress and egg laying. The task therefore consists in extracting a characteristic representation of the signals by assigning one or more semantic labels that indicates the state of the laying hens. Every vocalization can be assigned to multiple semantic labels, making this a multilabel classification problem. For example, a cold hen may utter a continuous vocalization because it is unhappy and in distress or in panic. Distress, fear and panic being distinct states which can arise from the same vocalization. This demonstrates the difficulty of the task addressed in this work in the sense that the objective is to categorize vocalizations that are acoustically very close for humans but which identify very distinct states for hens.

\section{Data collection}
\label{dataset}

Taking into account the diversity of breeds of hens and the context of breeding which strongly depends on the production environment, a major contribution of our work has been to build a database of vocalizations associated with certain behaviors of laying hens that we make available for future research in the field. We made recordings of hens from 10 different farms with beninese poultry farmers.
As they all come from modern commercial poultry farming, the laying hens from the target farms are under very specific veterinary treatment: deworming, anti-stress, regular vaccination. However, it happens that some hens in the group are exposed to physical disorders. 

For our work, only physically active laying hens were selected. 
Sound data were acquired continuously in Waveform Audio File Format (1 channel, 32-bit resolution, 16,000 Hz, recording at approximately 55s of each file). The recordings were made by using cameras positioned in the center about 2m from the hens and by considering the following vocalizations that are described in Table \ref{tab:description}: {\bf Alarm} - {\bf Food calls} - {\bf Gakel} - {\bf Egg laying} - {\bf Fear} - {\bf Distress} - {\bf Panic} - {\bf Lonely calls}. The recordings were then annotated by 4 poultry farmers who assigned labels to the different vocalizations. A total of 205 laying hens were recorded on all 8 vocalizations. Cross recordings were made so that several vocalizations are obtained with the same hen. During this experiment, 135 hours of data were recorded. The third column in Table \ref{tab:description} presents the number of signals obtained with the 8 vocalizations. The signal distribution in  Table \ref{tab:description}  indicates that the labelled dataset contains imbalanced sample sizes.

\begin{table}[htb]
\centering
\footnotesize
\caption{Description of the hen calls studied in this work.}
\begin{tabular}{@{}p{1.5cm}p{5.5cm}p{1cm}@{}}
\toprule
\textbf{Vocalization type} & \textbf{Description} & \textbf{Number of signals} \\ \midrule
Alarm & A normal sound of long duration (< 6s) with interruptions of less than (< 0.5s) often emitted by the hen to keep
everyone safe and aware of danger.  & 955 \\ 
Food calls & These are repetitive clucks emitted for either calls to feed or when the hens are hungry. the sounds are emitted in short sequences (<4s) for a long duration as long as food is not available to them & 1000 \\
Gakel calls & These are sounds of variable frequencies emitted to indicate
frustration with their environment. & 905 \\
Egg laying & The hens emit theses sounds (the signals of short sequences) when they are laying an egg or waiting to get into their favorite nest box. & 865 \\
Fear & These are sounds (high-pitched, repetitive, and fast-paced) that the hens make to express a fear behavior (running away from stressors, alerting of potential predators) & 1460 \\ 
Distress & These are sounds (very higher in pitch and repetitive) that indicate the hens are hungry (very similar to acoustic signals of the food calls) or annoyed by an unusual situation & 975 \\
Panic & Similar to a distress sound, this vocalization is also high-pitched and loud but more persistent with a long duration. & 975 \\
Lonely calls & The hens emit this call when they are alone and no company in their immediate environment. The signals are high-pitched and of very short duration (~2s) & 840 \\ \bottomrule
\end{tabular}
\label{tab:description}
\end{table}

\section{Our Multi-Label classification approach}
\label{approach}
We developed our study in two stages: an acoustic analysis of each vocalization sounds and the classification of the sound features based on recurrent neural network models and attention mechanism. 

\subsection{Acoustic features}

The acoustic analysis led to obtain the graphic representation of sound parameters from the popular features widely used in the literature. The following features were derived from the time and frequency domain. The time domain features are calculated around the syllables identified in a sound of the hen whereas the frequency domain features are directly from the raw signal.
These features are calculated by considering the syllable as acoustic unit. A syllable is represented by a vocal activity. These sounds (vocalizations) containing the syllables are interwoven between the parts of the silence and signal segments, which are not useful such as noise and non-vocal sound activities. The syllable
therefore has a variable size in a vocalization and depends on the presence or not of the silence markers in the signal. To extract the vocal syllables from the raw signal, we used the vocal activity detection algorithm proposed in \citep{MAHDAVIAN2020105100}.

The obtained syllables  are used to feed into the classification model. The time domain features considered in our work are: tempo ( number of syllables per second), energy, intensity, power and pitch.

The frequency domain features that we were considered are: formants (F1 to F4) and spectral energy characterized by the decomposition of the raw temporal signal into its constituent frequencies.

The formants and the spectral energy are augmented by the cepstral coefficients widely used in acoustic speech processing in order to capture more features and obtain a more robust representation of the hen vocalizations. These are the Mel Frequency Cepstral Coefficients (MFCC) and Linear Frequency Cepstral Coefficients (LFCC). They were calculated individually then merge afterwards into the same vector  to gather information across the whole spectrum. The MFCC is based on a nonlinear cepstral representation of the signal equally spaced on the Mel scale. This scale approximates the response of the human auditory system more closely than linear-spaced frequencies \citep{MAHDAVIAN2020105100, DU2021106221}. These cepstral coefficients are obtained by transforming from Hertz to Mel scale that is calculated as follows:

\begin{equation}
    mel = 2595log_{10}(\frac{f}{100} + 1)
\end{equation}

The LFCC coefficients are the result of taking the Discrete Fourier Transform (DFT) of the logarithm of the estimated spectrum of a signal \citep{app9194097}. These coefficients have been calculated using a linear filter bank, which has
better resolution in higher frequency regions. AS described in \citep{app9194097}, we based the calculation of the LFCC coefficients on a log energy filter bank applied to the discrete cosine transform as shown in \ref{lfcc}, where $j$ denotes the index of the LFCC giving the number of coefficients to compute, $X_i$ is the log-energy output from the ith filter and $B$, the number of triangular filters.

\begin{equation}
    \label{lfcc}
    L_j = \sum_{i=1}^{B} = X_i cos(j(i-\frac{1}{2})\pi/B)
\end{equation}

$40$ triangular filters are extracted as features for both MFCC and LFCC and $20ms$ a frame size. To extract more significant information from the vocalization signals, we merged the MFCC and LFCC coefficients to obtain a robust representation to feed the classification model. The merge consisted of creating a matrix by horizontally concatenating the LFCC and MFCC features. The features merged have been compared with each taken individually for comparison purposes.
\subsection{Archiecture of classification system.}

\begin{figure*}[btp]
\centering
  \includegraphics[scale=0.5]{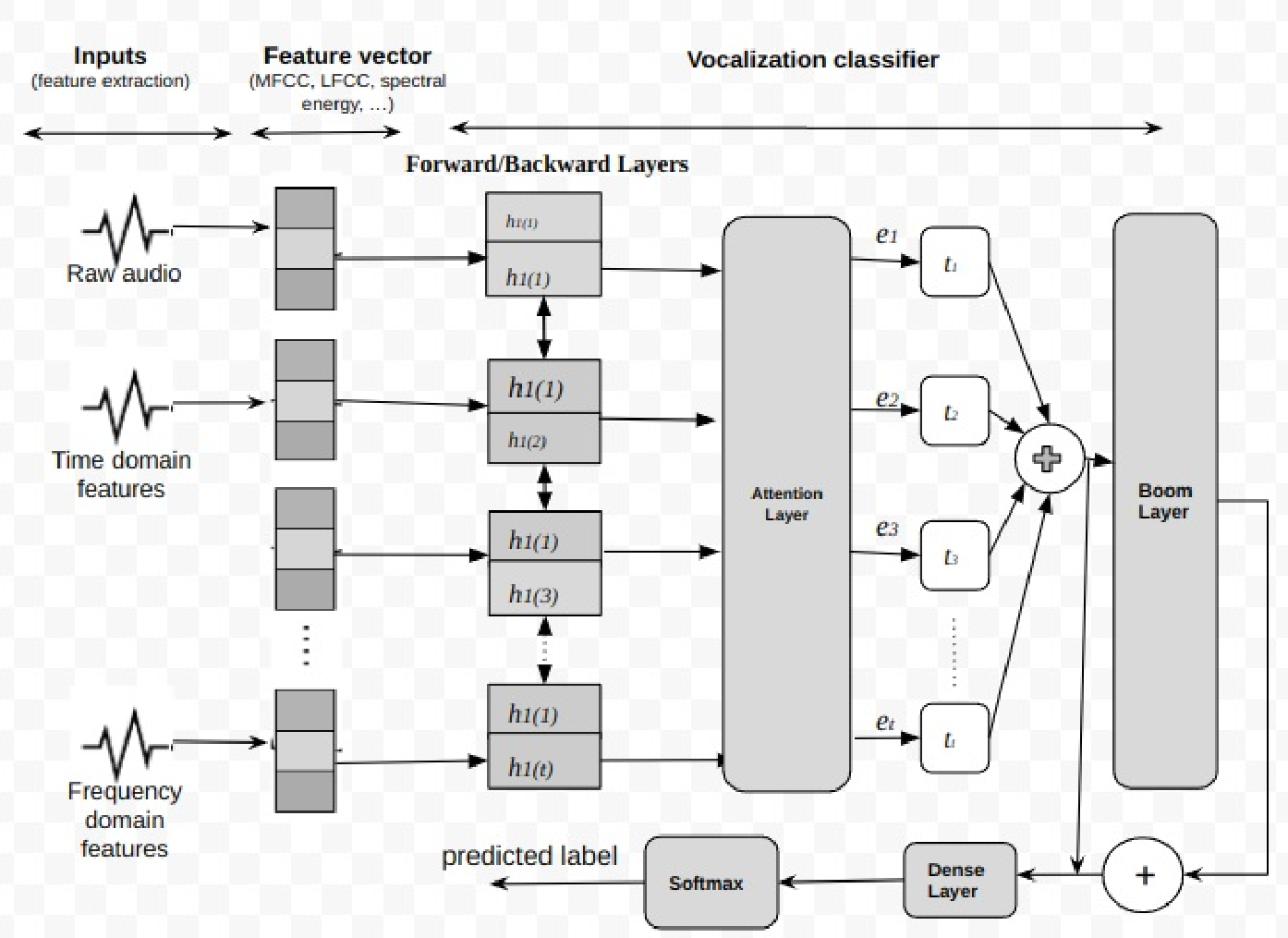}
  \caption{Architecture of the hen vocalization multi-label classification system.}
  \label{fig:arch}
\end{figure*}

We have implemented a classification approach based on recurrent neural networks (RNN) with attention mechanism that we have compared with standard and Bayesian-type approaches. Our proposed system is composed of three modules (Figure \ref{fig:arch}): input preprocessing, feature embedding, and the classifier. he input preprocessing step extracts temporal and frequency features from the vocalization sounds. The feature embedding module prepares the inputs for the model. The classifier module is composed of a trainable features layer, a stacked single head attention recurrent neural network, a Boom layer, and a softmax classifier.

Our proposed model differs from transformer-based approaches \cite{attention} in that we use a single attention head. This decision was inspired by \cite{DBLP:journals/corr/abs-1911-11423} and has resulted in improved performance in terms of the amount of useful information extracted. We chose not to complicate the model by adding multiple attention heads, as we were unsure of their benefits for the less complex signals that we worked with.
The attention mechanism is very computationally efficient since the only matrix multiplication performed during the process acts on the query  \cite{DBLP:journals/corr/abs-1911-11423}.

The boom layer, in our work, has been very useful in that it allows to control the explosion of the dimension of the vectors output from the attention layers: the vectors are projected to a higher dimension and projected back to original dimension.

Our model is loaded with large feed-forward layer steps at several points to encode the features from vocalization sounds across each feature dimension to produce dynamically vectors in hidden states. 


The produced vectors are used by attention layers  to weight the syllables contained in the vocalizations deferentially when aggregating them into a global representations. The attention mechanism led each syllable to contribute to the final vector used to map a vocalization to a semantic label. The Boom layer takes the final vector and uses the matrix multiplication obtained with  the attention weights to produce the final vector to  Dense and Softmax layers that output the predicted label. It is made up of connected layers
that at first blowing the input vector up to dimension 512 and then projecting it back down to dimension 8 the number of calls to predict.

\subsection{Objective}

Since it's basically impossible to have perfect predictions on close semantic labels (classes to predict) that share the acoustic similarities, we hypothesize that grouping in a semantic class, signals that are potentially semantically close would reduce the prediction errors of the model on certain labels. We have therefore created 4 master classes containing the primary labels. Table \ref{tab:master} presents the master classes and the semantic labels they contain. Given this previously mentioned difficulty, we used as an objective function a customized Binary Cross Entropy (cBCE) loss \citep{Jadon2020ASO} that we nested on each of the master classes. To do that, we firstly considered, in loss calculation, 4 times the logits of each master class which are masked and projected down onto the subclasses contained in each master class. This consisted of taking the top 2 logists from each master class, mixing them to compute the cBCE which is actually the weighted binary cross-entropy loss. Then and based on Equation \ref{fig:bce}, two losses are calculated: one on the logits of the subclasses and the other on the logits of the master classes. In Equation \ref{fig:bce}, $\alpha$  is used to tune the false negative and the false positive predictions and $\hat{y}$ is the predicted value by the prediction model. Finally, we take a weighted sum of the two losses for indicating to model which priority it should give to master class loss over the subclass loss.

\begin{table}[!htpb]
\centering
\caption{Master classes.}
\begin{tabular}{lc}
\toprule
\textbf{Master class} & \textbf{Label}  \\ \midrule
Food calls & Food calls - Distress - Panic \\ 
Egg laying & Egg laying \\
Fear & Fear - Alarm - Gakel calls \\
Lonely calls & Lonely calls \\
\bottomrule
\end{tabular}

\label{tab:master}
\end{table}

\begin{equation}
    L(y, \hat{y}) = - (\alpha*y log(\hat{y}) + (1 - y)log(1 - \hat{y}))
    \label{fig:bce}
\end{equation}
\section{Results and discussion}
\label{results}

we have set up and conducted different experimental protocols based on combinations of the different features in order to produce a more accurate model for the recognition of vocalizations. For each experimental setup, we used as input data the combination of features built and evaluated the contribution of the acoustic information provided by each of them. 

\subsection{Experiment setup}

Different configurations have been implemented to test the efficiency and accuracy of the approach we propose. First, we compared our approach implemented on different feature combination configurations to the techniques commonly used in the state of the art for animal behavior recognition task such as: Decision Tree (DT) \cite{CHELOTTI201883}, Support Vector Machine \cite{bishop2017sound, Lee2015StressDA}, Random Forest (RF) \cite{app9194097} and Gaussian Mixture Model \citep{DU2021106221}. 
In order to consolidate our hypothesis of using a single model for all studied vocalizations and not training multiple models per vocalization, we compared our model's multi-label approach, with nested loss calculation, to a cascading binary classification approach followed by a majority vote. This evaluation served to determine which approach was most effective in achieving our desired outcome.

We used all of the $\sim8000$ samples for model evaluation with several splits to build the training, validation and testing sets. First, we set aside $\sim20\%$ of the data to obtain the testing set, ensuring that all labels are represented at almost equal rates so as not to penalize the models in their evaluation. We used $10$-fold cross-validation for training and validation with a ratio $90/10$ for splitting the dataset. The models were trained on different training data obtained from the random generation at different places. With  validation set,  we obtained  the  best-performing  parameters (Table \ref{tab:parameters})  that  are used  for  the  evaluation  of our model. 

\begin{table}[!htpb]
\centering
\footnotesize
\caption{Hyperparameter values}
\begin{tabular}{@{}p{0.5cm}p{0.5cm}p{0.5cm}p{1cm}p{0.8cm}p{0.8cm}p{0.5cm}@{}}
\toprule
\textbf{Hidden size} & \textbf{Boom dim} & \textbf{Batch size} & \textbf{cBCE ratio} & \textbf{Dropout} & \textbf{learning rate} & \textbf{Optimizer} \\ \midrule
1024 & 512 & 16 & 0.1 & 0.2 & 0.001 & Adamdelta \\ \bottomrule
\end{tabular}

\label{tab:parameters}
\end{table}

Experiments on binary classifiers are done by training a model by vocalization, which yields 8 different models for the same input data configuration. On all 8 models, we applied a decision-making strategy based on both the majority voting of the models and a weighted linear interpolation of the classification model results. This made it possible to obtain the better performances of each of the models. The label predicted by a binary classifier is the one that obtained the most votes from the 8 models. The
final label is given by the binary classifier having predicted a label with the highest probability. 

Since we are in a multilabel setup, we used as evaluation metric the sample-average $F1$ score by computing the score of the predictions of the sample and averaging that over all the samples. Also, F1 was chosen for this work because of the imbalance of call classes present in our training dataset.

\subsection{Using time domain features}

Our first experiments are made using the features calculated only in the time domain. The input vector used to train the different models is formed from the concatenation of Tempo, Energy, Intensity, Power and Pitch. Table \ref{tab:multime} and Table \ref{tab:binary} present respectively the results of multi-output models and binary models on testing set.
Table \ref{tab:multime} shows that the GMM model achieved the highest F1-score of $33.15$, followed by the SVM model with $32.95$, indicating that they are better suited for time domain features. However, our RNN-based model achieved a significantly lower F1-score of $22.85$, suggesting that time domain features may not provide sufficient information for capturing the acoustic representations of vocalizations. The same observation can be made in
Table \ref{tab:binary} that shows the results of binary classifiers on the test data for different types of vocalizations. Based on the performance provided by time domain features, the SVM model performed the best using time domain features, achieving a high accuracy in predicting panic and fear vocalizations. The GMM model performed the best overall, achieving the highest accuracy in predicting egg laying, panic, fear and alarm calls. 

On average, when using majority voting, the SVM and GMM models demonstrate superior performance in classifying all vocalizations compared to the other models. Our model performs relatively poorly on both types of classification, with an average increase of $10$ points in binary classification compared to multi-output classification.

\begin{table}[!htpb]
\centering
\footnotesize
\caption{Evaluation of multi-label models using time domain features on Test data }
\begin{tabular}{@{}ll@{}}
\toprule
\textbf{Model} & \textbf{F1-score} \\ \midrule
Decision Tree & 25.12 \\
Support Vector Machine & 32.95 \\
Random Forest & 22.65 \\
GMM & 33.15 \\
Our model & 22.85 \\ \bottomrule
\end{tabular}

\label{tab:multime}
\end{table}

\begin{table}[!hb]
\caption{Evaluation of binary classifiers on test data.}
\footnotesize
\begin{tabular}{@{}cccccccccc@{}}
\toprule
 \multirow{2}{*}{\textbf{Model}} & \multicolumn{3}{c}{\textbf{Food calls}} & \textbf{Egg Laying} & \multicolumn{3}{c}{\textbf{Fear}} & \textbf{Lonely calls} & \multirow{2}{*}{\textbf{Majority voting}}  \\ \cmidrule{2-9}
  & Food calls & Distress & Panic & Egg Laying & Fear & Alarm & Gakel calls & Lonely calls \\ \midrule
\multicolumn{10}{c}{\textbf{Using time domain features}} \\ \midrule
Decision Tree & 41.85 & 42.33 & 37.55 & 35.10 & 46,75 & 40.60 & 40.15 & 39.87 & 45.20 \\
SVM & 43.15 & 44.13 & 49.65 & 46.10 & 46.70 & 42.45 & 41.75 & 39.88 & 47.12 \\
Random Forest & 39.5 & 39.63 & 35.65 & 33.80 & 30.69 & 38.60 & 37.75 & 29.87 & 30.80 \\
GMM & 45.01  & 46.01 & 49.88 & 51.15 & 49.75 & 48.88 & 43.12 & 41.15 & 47.75 \\
Our model & 33.15 & 35.33 & 32.10 & 28.95 & 34,35 & 30.80 & 30.75 & 29.47 & 31.95 \\
\midrule
\multicolumn{10}{c}{\textbf{Using frequency domain features}} \\ \midrule
Decision Tree & 51.35 & 61.45 & 47.65 & 45.10 & 56.45 & 52.20 & 50.75 & 41.87 & 50.20 \\
SVM & 63.25 & 65.33 & 69.65 & 56.10 & 51.75 & 52.82 & 50.65 & 50.88 & 57.65 \\
Random Forest & 59.55 & 59.63 & 55.65 & 54.85 & 57.09 & 53.40 & 54.75 & 49.67 & 54.88 \\
GMM & 66.15 & 66.22 & 66.95 & 55.84 & 56.85 & 53.05 & 55.55 & 51.15 & 56.55 \\ 
Our model & 73.05 & 65.33 & 69.95 & 60.75 & 54,75 & 58.88 & 57.45 & 57.47 & 60.05 \\
\midrule
\multicolumn{10}{c}{\textbf{Using Time-Frequency domain features}} \\ \midrule 
Decision Tree & 60.75 & 65.55 & 60.65 & 55.40 & 62.45 & 62.20 & 60.74 & 54.77 & 57.22 \\
SVM & 68.75 & 72.30 & 76.25 & 66.40 & 56.25 & 58.52 & 55.75 & 59.78 & 62.25 \\
Random Forest & 62.55 & 63.63 & 65.35 & 58.15 & 63.29 & 65.40 & 59.75 & 55.07 & 58.35 \\
GMM & 70.25 & 70.15 & 77.85 & 66.45 & 64.25 & 59.75 & 60.15 & 58.58 & 65.12 \\
Our model & 78.05 & 75.35 & 81.95 & 70.15 & 62.77 & 63.88 & 60.55 & 60.47 & 75.65 \\
\bottomrule
\end{tabular}

\label{tab:binary}
\end{table}

\subsection{Using frequency domain features}

Four combined features were considered in these experiments for training and testing the different models such as Formants+Spectral energy, MFCCs, LFCCs, MFCCs+LFCCs. 
These types of combination can be computationally intensive yet provide valuable information on formants and spectral energy, hence motivating their incorporation into feature combinations. Our choice to combine features is influenced by the latest advances in research especially by the work in \citep{DU2021106221}, where the authors demonstrated the effectiveness of combining tristimulus values with formants to classify the different call types of laying hens.
Table \ref{tab:freq} shows the evaluation evaluation of multi-label models using frequency domain features on the test data. The MFCCs+LFCCs feature set achieved the highest F1-score for all models, with our model achieving the highest score of $79.85$, which is significantly higher than the other models. This suggests that the combination of MFCCs and LFCCs provides the best acoustic representation of vocalizations for our RNN-based model to accurately classify different types of vocalizations. 

Additionally, it is noteworthy that using only {\it Formants+Spectral energy}, the models achieve twice the accuracy of the time feature models. These observations are consistent with those in Table  \ref{tab:binary}, where our model using majority voting achieved significant performance with an F1-score of $60.05$ in binary classification, which approaches acceptable accuracies for a vocalization recognition system of laying hens.

\begin{table}[!htpb]
\centering
\footnotesize
\caption{Evaluation of multi-label models using frequency domain features on Test data }
\begin{tabular}{@{}lp{1cm}lp{1cm}p{1cm}p{1cm}@{}}
\toprule
\textbf{Features} & \textbf{Decision Tree} & \textbf{SVM} & \textbf{Random Forest} & \textbf{GMM} & \textbf{Our model} \\ \midrule
Formants+Spectral energy & 45.79 & 57.85 & 50.75 & 57.95 & 59.95 \\
MFCCs & 52.05 & 60.15 & 53.65 & 62.62 & 72.63 \\
LFCCs & 52 & 59.15 & 53.66 & 61.55 & 72.61 \\
MFCCs+LFCCs & 54.15 & 66.67 & 55.67 & 68.45 & 79.85 \\ \bottomrule
\end{tabular}

\label{tab:freq}
\end{table}

\subsection{Using Time-frequency domain features}

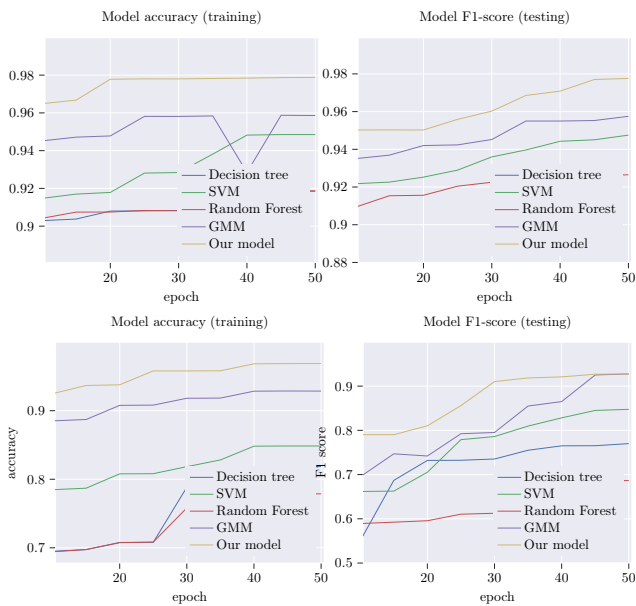
\begin{figure}[htb]
\centering
\resizebox{8.5cm}{!}{%
\begin{tikzpicture}

\definecolor{color0}{rgb}{0.917647058823529,0.917647058823529,0.949019607843137}
\definecolor{color1}{rgb}{0.298039215686275,0.447058823529412,0.690196078431373}
\definecolor{color2}{rgb}{0.333333333333333,0.658823529411765,0.407843137254902}
\definecolor{color3}{rgb}{0.768627450980392,0.305882352941176,0.32156862745098}
\definecolor{color4}{rgb}{0.505882352941176,0.447058823529412,0.698039215686274}
\definecolor{color5}{rgb}{0.8,0.725490196078431,0.454901960784314}

\begin{groupplot}[group style={group size=2 by 1}]
\nextgroupplot[
axis background/.style={fill=color0},
axis line style={white},
legend cell align={left},
legend style={at={(0.97,0.03)}, anchor=south east, draw=none, fill=color0},
tick align=outside,
tick pos=left,
title={Model accuracy (training)},
x grid style={white},
xlabel={epoch},
xmajorgrids,
xmin=10.45, xmax=50.45,
xtick style={color=white!15.0!black},
y grid style={white},
ymajorgrids,
ymin=0.880800427178351, ymax=0.999792724310862,
ytick style={color=white!15.0!black}
]
\addplot [line width=0.7000000000000001pt, color1]
table {%
5 0.901353863909546
10 0.902909922539072
15 0.90376248785222
20 0.90797778557304821
25 0.908093407450595
30 0.908268324198247
35 0.908688124392611
40 0.918355782572072
45 0.918408257596368
50 0.918513207613682
};
\addlegendentry{Decision tree}
\addplot [line width=0.7000000000000001pt, color2]
table {%
5 0.910671686977695
10 0.914735005478645
15 0.916973940265623
20 0.917866015678647
25 0.92811089912536
30 0.928390765921603
35 0.938163374149656
40 0.948233340848716
45 0.948513207644959
50 0.948513207644959
};
\addlegendentry{SVM}
\addplot [line width=0.7000000000000001pt, color3]
table {%
5 0.909709644839545
10 0.904157780284375
15 0.907446215484283
20 0.907463707159048
25 0.908320799222542
30 0.908163374149656
35 0.908478224295429
40 0.91814588247489
45 0.918670632717846
50 0.918688124392611
};
\addlegendentry{Random Forest}
\addplot [line width=0.7000000000000001pt, color4]
table {%
5 0.93880007704801
10 0.945172297764809
15 0.947148857013275
20 0.947796048979587
25 0.95814588247489
30 0.958128390800125
35 0.958390765869474
40 0.928478224295429
45 0.958740599416907
50 0.958600666018785
};
\addlegendentry{GMM}
\addplot [line width=0.7000000000000001pt, color5]
table {%
5 0.960269378353838
10 0.964874939189542
15 0.966781531738947
20 0.977813540654352
25 0.978023440647275
30 0.978005949076769
35 0.978233340848716
40 0.978408257596368
45 0.978688124392611
50 0.978793074441202
};
\addlegendentry{Our model}

\nextgroupplot[
axis background/.style={fill=color0},
axis line style={white},
legend cell align={left},
legend style={at={(0.97,0.03)}, anchor=south east, draw=none, fill=color0},
tick align=outside,
tick pos=left,
title={Model F1-score (testing)},
x grid style={white},
xlabel={epoch},
xmajorgrids,
xmin=10.45, xmax=50.45,
xtick style={color=white!15.0!black},
y grid style={white},
ymajorgrids,
ymin=0.880, ymax=0.999132443281716,
ytick style={color=white!15.0!black}
]
\addplot [line width=0.7000000000000001pt, color1]
table {%
5 0.83015542
10 0.8303625
15 0.8318954
20 0.8320000
25 0.8323522
30 0.8352007
35 0.8550002
40 0.865001
45 0.865263
50 0.87
};
\addlegendentry{Decision tree}
\addplot [line width=0.7000000000000001pt, color2]
table {%
5 0.92023254
10 0.92174585
15 0.92259632
20 0.92524856
25 0.92896528
30 0.93596852
35 0.93965968
40 0.94425412
45 0.94504582
50 0.94752065
};
\addlegendentry{SVM}
\addplot [line width=0.7000000000000001pt, color3]
table {%
5 0.90026324
10 0.90926325
15 0.91536256
20 0.91562358
25 0.92045289
30 0.92252863
35 0.92256391
40 0.92504285
45 0.92602596
50 0.926502145
};
\addlegendentry{Random Forest}
\addplot [line width=0.7000000000000001pt, color4]
table {%
5 0.9255542
10 0.93503625
15 0.9368954
20 0.9420000
25 0.9423522
30 0.9452007
35 0.9550002
40 0.955001
45 0.955263
50 0.9575
};
\addlegendentry{GMM}
\addplot [line width=0.7000000000000001pt, color5]
table {%
5 0.950145823
10 0.95025632
15 0.95026325
20 0.95025336
25 0.95589622
30 0.96025638
35 0.96856325
40 0.97085693
45 0.97701258
50 0.97758962
};
\addlegendentry{Our model}
\end{groupplot}

\end{tikzpicture}
}
\resizebox{8.5cm}{!}{%
\begin{tikzpicture}

\definecolor{color0}{rgb}{0.917647058823529,0.917647058823529,0.949019607843137}
\definecolor{color1}{rgb}{0.298039215686275,0.447058823529412,0.690196078431373}
\definecolor{color2}{rgb}{0.333333333333333,0.658823529411765,0.407843137254902}
\definecolor{color3}{rgb}{0.768627450980392,0.305882352941176,0.32156862745098}
\definecolor{color4}{rgb}{0.505882352941176,0.447058823529412,0.698039215686274}
\definecolor{color5}{rgb}{0.8,0.725490196078431,0.454901960784314}

\begin{groupplot}[group style={group size=3 by 1}]
\nextgroupplot[
axis background/.style={fill=color0},
axis line style={white},
legend cell align={left},
legend style={at={(0.97,0.03)}, anchor=south east, draw=none, fill=color0},
tick align=outside,
tick pos=left,
title={Model accuracy (training)},
x grid style={white},
xlabel={epoch},
xmajorgrids,
xmin=10.45, xmax=50.45,
xtick style={color=white!15.0!black},
y grid style={white},
ylabel={accuracy},
ymajorgrids,
ymin=0.677800427178351, ymax=0.999792724310862,
ytick style={color=white!15.0!black}
]
\addplot [line width=0.7000000000000001pt, color1]
table {%
5 0.681353863909546
10 0.694909922539072
15 0.697376248785222
20 0.707778557304821
25 0.708093407450595
30 0.788268324198247
35 0.798688124392611
40 0.818355782572072
45 0.818408257596368
50 0.818513207613682
};
\addlegendentry{Decision tree}
\addplot [line width=0.7000000000000001pt, color2]
table {%
5 0.750671686977695
10 0.784735005478645
15 0.786973940265623
20 0.807866015678647
25 0.80811089912536
30 0.818390765921603
35 0.828163374149656
40 0.848233340848716
45 0.848513207644959
50 0.848513207644959
};
\addlegendentry{SVM}
\addplot [line width=0.7000000000000001pt, color3]
table {%
5 0.689709644839545
10 0.694157780284375
15 0.697446215484283
20 0.707463707159048
25 0.708320799222542
30 0.758163374149656
35 0.758478224295429
40 0.77814588247489
45 0.778670632717846
50 0.778688124392611
};
\addlegendentry{Random Forest}
\addplot [line width=0.7000000000000001pt, color4]
table {%
5 0.82880007704801
10 0.885172297764809
15 0.887148857013275
20 0.907796048979587
25 0.90814588247489
30 0.918128390800125
35 0.918390765869474
40 0.928478224295429
45 0.928740599416907
50 0.928600666018785
};
\addlegendentry{GMM}
\addplot [line width=0.7000000000000001pt, color5]
table {%
5 0.920269378353838
10 0.924874939189542
15 0.936781531738947
20 0.937813540654352
25 0.958023440647275
30 0.958005949076769
35 0.958233340848716
40 0.968408257596368
45 0.968688124392611
50 0.968793074441202
};
\addlegendentry{Our model}

\nextgroupplot[
axis background/.style={fill=color0},
axis line style={white},
legend cell align={left},
legend style={at={(0.97,0.03)}, anchor=south east, draw=none, fill=color0},
tick align=outside,
tick pos=left,
title={Model F1-score (testing)},
x grid style={white},
xlabel={epoch},
xmajorgrids,
xmin=10.45, xmax=50.45,
xtick style={color=white!15.0!black},
y grid style={white},
ylabel={F1 score},
ymajorgrids,
ymin=0.50, ymax=0.999132443281716,
ytick style={color=white!15.0!black}
]
\addplot [line width=0.7000000000000001pt, color1]
table {%
5 0.5055542
10 0.5503625
15 0.6868954
20 0.7320000
25 0.7323522
30 0.7352007
35 0.7550002
40 0.765001
45 0.765263
50 0.77
};
\addlegendentry{Decision tree}
\addplot [line width=0.7000000000000001pt, color2]
table {%
5 0.66023254
10 0.66174585
15 0.66259632
20 0.70524856
25 0.77896528
30 0.78596852
35 0.80965968
40 0.82825412
45 0.84504582
50 0.84752065
};
\addlegendentry{SVM}
\addplot [line width=0.7000000000000001pt, color3]
table {%
5 0.55026324
10 0.58926325
15 0.59236256
20 0.59562358
25 0.61045289
30 0.61252863
35 0.67256391
40 0.68504285
45 0.68602596
50 0.686502145
};
\addlegendentry{Random Forest}
\addplot [line width=0.7000000000000001pt, color4]
table {%
5 0.6855542
10 0.69503625
15 0.7468954
20 0.7420000
25 0.7923522
30 0.7952007
35 0.8550002
40 0.865001
45 0.925263
50 0.9275
};
\addlegendentry{GMM}
\addplot [line width=0.7000000000000001pt, color5]
table {%
5 0.720145823
10 0.79025632
15 0.79026325
20 0.81025336
25 0.85589622
30 0.91025638
35 0.91856325
40 0.92085693
45 0.92701258
50 0.92758962
};
\addlegendentry{Our model}
\end{groupplot}

\end{tikzpicture}
}
\caption{Accuracies and F1 scores of models}
\label{fig:acc_loss_fig1}
\end{figure}

In this experiment, we used three inputs to feed the models in parallel. Time features, Formants+Spectral energy and MFCCs+LFCCs are used separately with different acoustic representations that feed the LSTM cells that propagate information so that it is shared efficiently across layers of neurons (see Figure \ref{fig:arch}). Table \ref{tab:final} shows the results of this experiment.
It can be observed that the GMM and our model outperform the other models with F1-scores of $89.55$ and $92.75$, respectively. The SVM model also achieves a high F1-score of 84.65. In table \ref{tab:binary}, using time-domain features, our model again shows the highest F1-score for all categories, with an outstanding F1-score of $81.95$ for the "Panic" category. This suggests that the frequency information augmented by the time domain features enriched the models and made them more accurate. In Figure \ref{fig:acc_loss_fig1}, we can observed that our model was the most accurate with $F1=92.75$ compared to other models ( GMM=$89.55$, SVM=$84.65$, DT=$77.15$ and Random Forest=$68.65$ ) which, with this feature enrichment strategy, outperformed their performance obtained with MFCCs+LFCCs as inputs.

In order to contextualize our approach, which relies on RNNs and feature combination strategies, we conducted a comparative analysis between our model and the GMM model explored in \citep{DU2021106221} under similar settings. To adapt to our own dataset, which differs from that used by the authors, we implemented their feature calculation and extraction strategies. To ensure consistency and a fair basis for comparison, we chose accuracy as the metric for evaluating the models, rather than relying on the results reported by the authors.

The accuracies obtained are listed in Table \ref{tab:gmm}, indicating that our model outperforms the GMM from \citep{DU2021106221} for all feature combinations. The authors observed a significant improvement in performance by introducing Tristimulus-formant as a new feature and combining it with MFCCs. We were able to replicate this result on our own dataset using Tristimulus-formant. However, our proposed model and the combination MFCCs+LFCCs achieved even higher performance compared to the feature combinations they proposed. 

To delve deeper, we crafted the confusion matrices by evaluating our model (see Table \ref{tab:conf1}) and GMM (see Table \ref{tab:conf2}) on the validation set, taking into account their performance on other implemented models. Our objective is to compare these two matrices to detect disparities in precision and recall for each class. To do so, we utilized the master classes of the vocalizations of laying hens. From the precision and recall values shown in Table \ref{tab:conf1}, we can infer that our model outperforms GMM in Table \ref{tab:conf2}, with higher precision and recall values for all classes. This suggests that, progressing hierarchically in the master classes, our model is better equipped to accurately identify laying hen calls than the GMM.

\begin{table}[!ht]
\centering
\footnotesize
\caption{Evaluation of multi-label models using the three inputs data.}
\begin{tabular}{@{}ll@{}}
\toprule
\textbf{Model} & \textbf{F1-score} \\ \midrule
Decision Tree & 77.15 \\
Support Vector Machine & 84.65 \\
Random Forest & 68.65 \\
GMM & 89.55 \\
Our model & 92.75 \\ \bottomrule
\end{tabular}

\label{tab:final}
\end{table}


\begin{table}[!ht]
\centering
\footnotesize
\caption{Accuracy rate of our RNN-based model and the GMM calculated using the different combinations explored in \citep{DU2021106221}. TF meaning "tristimulus-forming", is the novel feature introduced by the authors to train their model of recognition of the laying hen calls.}
\begin{tabular}{@{}lllll@{}}
\toprule
\textbf{Model} & \textbf{MFCCs} & \textbf{Formants+TF} & \textbf{MFCCs+TF} & \textbf{MFCCs+LFCCs} \\ \midrule
GMM \citep{DU2021106221} & 90.79 & 68.55 & 70.12 & 92.80 \\
Our model & 93.45 & 60.15 & 82.50 & 96.25 \\ \bottomrule
\end{tabular}

\label{tab:gmm}
\end{table}

\begin{table}[!htpb]
\centering
\footnotesize
\caption{Confusion matrix from our model evaluation on a validation set using the three input data.}
\begin{tabular}{@{}lp{1.5cm}lp{1.5cm}p{1.5cm}@{}}
\toprule
  & \textbf{Food calls} & \textbf{Egg laying} & \textbf{Fear} & \textbf{Lonely calls} \\ \midrule
Food calls & 490 & 0 & 20 & 0 \\
Egg laying & 0 & 177 & 0 & 0  \\
Fear & 70 & 30 & 570 & 20  \\
Lonely calls & 10 & 0 & 40 & 216  \\ \bottomrule
Precision (\%) & 86  & 86 & 90 & 92  \\
Recall (\%) & 96 & 1 & 83 & 81 \\ \bottomrule 
\end{tabular}
\label{tab:conf1}
\end{table}

\begin{table}[!htpb]
\centering
\footnotesize
\caption{Confusion matrix from GMM evaluation on a validation set using the three input data.}
\begin{tabular}{@{}lp{1.5cm}lp{1.5cm}p{1.5cm}@{}}
\toprule
  & \textbf{Food calls} & \textbf{Egg laying} & \textbf{Fear} & \textbf{Lonely calls} \\ \midrule
Food calls & 390 & 77 & 70 & 0 \\
Egg laying & 10 & 100 & 0 & 0  \\
Fear & 100 & 30 & 470 & 46  \\
Lonely calls & 70 & 0 & 90 & 200  \\ \bottomrule
Precision (\%) & 68  & 48 & 75 & 81  \\
Recall (\%) & 73 & 91 & 76 & 56 \\ \bottomrule 
\end{tabular}
\label{tab:conf2}
\end{table}

\subsection{Discussion}

As shown in Figure \ref{fig:acc_loss_fig1} and in Tables \ref{tab:final}, \ref{tab:conf1}, and \ref{tab:conf2}, our proposed approach allowed to train the best and optimal model for recognizing laying hen calls which share the semantic similarities. Although the combination of MFCCs and LFCCs allowed to obtain better scores ($F1=79.85$), their enrichment by the time domain parameters and the formants using different input channels was the best strategy of our approach of behavior recognition of laying hens. It achieves a significant gain of $13\%$ using our model and an average of $18\%$ for other models. The disadvantage of our approach is that it is very computationally expensive during training and inference. Indeed, our model takes longer to train and predict while the methods used in most state-of-the-art works \citep{DU2021106221, app9194097} and used in this work for comparison purposes take less time to train with no less negligible performance ($F1=89.55$ for GMM and $F1=84.65$). 
Our model is, individually, more accurate on the Food call classes ($F1=81.95$ for Panic and $F1=78.05$ for Food calls) and less accurate on Lonely ($F1=60.47$) and Gakel ($F1=60.55$) calls. On the other hand, on the master classes, our model produces a global performance on our data compared to the approaches proposed in the state of the art. It is worth noting that
it is fair to compare the results of our approach to state-of-the-art methods proposed \citep{DU2021106221, app9194097}  on datasets they used for reasons of accessibility and also, we have not studied the same types of laying hen calls whose characteristics may vary according to breed, environment and diet. 

Our proposed approach in this work achieved the best results for recognizing laying hen calls that share semantic similarities, as illustrated in  Figure \ref{fig:acc_loss_fig1} and in Tables \ref{tab:final}, \ref{tab:conf1}, and \ref{tab:conf2}. By enriching the MFCCs and LFCCs with time domain parameters and formants from different input channels, our approach obtained an F1 score of $79.85$, which is $13\%$ better than the other models and $18\%$ better on average. However, this improvement comes at the cost of increased computational complexity during training and inference. While our model is more accurate on Food call classes (with F1 scores of $81.95$ and $78.05$ for Panic and Food calls, respectively), it is less accurate on Lonely (F1 score of $60.47$) and Gakel (F1 score of $60.55$) calls. Nevertheless, our model outperforms state-of-the-art approaches \citep{DU2021106221, app9194097} on master classes using our data. It should be noted that comparing our results to those of state-of-the-art methods on their respective datasets may prove challenging due to accessibility issues. Additionally, our study focused on a specific type of laying hen call, which may vary in characteristics based on factors such as breed, environment, and diet. Therefore, caution should be taken when generalizing our findings to other datasets and types of laying hen calls.

\section{Conclusion}

In this paper, we proposed a recognition system that makes it possible to detect and recognize earlier states related to laying hen  behaviour using their vocalization sounds. We described a multi-label classification approach based on intelligent combination of time and frequency domain features. Our strategy has been to leverage from the acoustic information generated by MFCCs and LFCCs features augmented by formants and spectral energy to build the rich and robust acoustic representations. These representations allowed us to feed our multi-label classification model which combines recurrent neural networks and the attention mechanism. With MFCCs+LFCCs augmented by the formants and time domain features, our system significantly improves its performance and obtained the overall performance.  The final system is more efficient on food
calls and more precisely on panic call despite having fewer signals than the Food and fear calls. This can demonstrate that the impact of data imbalance in modeling of a rich and robust representation of laying hen calls is relatively low in our study context. That finding remains to be proven in cases where each call would have the same number of signals and with more data to collect. This is a limitation of our experiments. We are currently preparing equipment and sensors to collect more audio data and add images from videos to build a multimodal monitoring system for laying hens in Beninese poultry farms.

\section{Availability of data and material}

The data that support the findings of this study are available from the corresponding author in the controlled access repository\footnote{Location: \url{https://github.com/laleye/ScreamAnalysis}}. Requests for material should also be addressed to the corresponding authors.

\section{Conflicts of interest}

All authors certify that they have no affiliations with or involvement in any organization or entity with any financial interest or non-financial interest in the subject matter or materials discussed in this manuscript.


\bibliography{sn-bibliography}


\end{document}